\documentclass[conference]{IEEEtran}
\IEEEoverridecommandlockouts
\usepackage{cite}
\usepackage{svg}
\usepackage{amsmath,amssymb,amsfonts}
\usepackage{algorithmic}
\usepackage{graphicx}
\usepackage{textcomp}
\usepackage{xcolor}
\usepackage{booktabs} 
\usepackage{graphicx}
\usepackage{xcolor}
\usepackage{xspace}
\usepackage[normalem]{ulem}
\useunder{\uline}{\ul}{}
\usepackage{lscape}
\usepackage{rotating}
\usepackage{supertabular}
\usepackage{longtable}
\usepackage{enumerate}
\usepackage{svg}
\usepackage{amsmath}
\usepackage{colortbl}
\usepackage{subcaption}
\usepackage{hhline}
\usepackage{makecell}
\usepackage{caption}
\usepackage{hyperref}
\usepackage{enumitem}
\usepackage{tcolorbox}
\usepackage{tabularx}
\usepackage{lipsum}
\usepackage{listings}
\usepackage{tabularx,ragged2e}
\usepackage{verbatim}
\usepackage{fancyvrb}
\usepackage{balance}
\usepackage{multirow}

\def\BibTeX{{\rm B\kern-.05em{\sc i\kern-.025em b}\kern-.08em
    T\kern-.1667em\lower.7ex\hbox{E}\kern-.125emX}}
\begin{document}

\title{Practices and Challenges of Using GitHub Copilot: An Empirical Study
\thanks{This work is funded by the NSFC with Grant No. 62172311 and the Special Fund of Hubei Luojia Laboratory.}
}




\author{
    \IEEEauthorblockN{Beiqi Zhang\IEEEauthorrefmark{2}\IEEEauthorrefmark{3}, Peng Liang\IEEEauthorrefmark{2}\IEEEauthorrefmark{3}\IEEEauthorrefmark{1}, Xiyu Zhou\IEEEauthorrefmark{2}\IEEEauthorrefmark{3}, Aakash Ahmad\IEEEauthorrefmark{4}, Muhammad Waseem\IEEEauthorrefmark{2}\IEEEauthorrefmark{3}\thanks{DOI reference number: 10.18293/SEKE2023-077}}
    \IEEEauthorblockA{
    \IEEEauthorrefmark{2}School of Computer Science, Wuhan University, Wuhan, China\\
    \IEEEauthorrefmark{3}Hubei Luojia Laboratory, Wuhan, China\\
    \IEEEauthorrefmark{4}School of Computing and Communications, Lancaster University Leipzig, Leipzig, Germany\\
    \{zhangbeiqi, liangp, xiyuzhou, m.waseem\}whu.edu.cn, a.ahmad13@lancaster.ac.uk}
}

\maketitle

\begin{abstract}
With the advances in machine learning, there is a growing interest in AI-enabled tools for autocompleting source code. GitHub Copilot, also referred to as the ``AI Pair Programmer'', has been trained on billions of lines of open source GitHub code, and is one of such tools that has been increasingly used since its launch in June 2021. However, little effort has been devoted to understanding the practices and challenges of using Copilot in programming with auto-completed source code. To this end, we conducted an empirical study by collecting and analyzing the data from Stack Overflow (SO) and GitHub Discussions. More specifically, we searched and manually collected 169 SO posts and 655 GitHub discussions related to the usage of Copilot. We identified the programming languages, IDEs, technologies used with Copilot, functions implemented, benefits, limitations, and challenges when using Copilot. The results show that when practitioners use Copilot: (1) The major programming languages used with Copilot are \textit{JavaScript} and \textit{Python}, (2) the main IDE used with Copilot is \textit{Visual Studio Code}, (3) the most common used technology with Copilot is \textit{Node.js}, (4) the leading function implemented by Copilot is \textit{data processing}, (5) the significant benefit of using Copilot is \textit{useful code generation}, and (6) the main limitation encountered by practitioners when using Copilot is \textit{difficulty of integration}. Our results suggest that using Copilot is like a double-edged sword, which requires developers to carefully consider various aspects when deciding whether or not to use it. Our study provides empirically grounded foundations and basis for future research on the role of Copilot as an AI pair programmer in software development.
\end{abstract}


\section{Introduction}
\label{sec:introduction}

Large Language Models (LLMs) and Machine Learning (ML) for autocompleting source code are becoming more and more popular in software development. LLMs nowadays incorporate powerful capabilities for Natural Language Processing (NLP) \cite{austin2021program}, and ML approaches have been widely applied to source code text in a variety of new tools to support software development \cite{allamanis2018survey}, which makes it possible to use LLMs to synthesize code in general-purpose languages \cite{austin2021program}. Recently, NLP-based code generation tools have come into the limelight, with generative pre-trained language models trained on large amounts of code in an attempt to provide reasonable auto-completion of the source code when programmers write code \cite{hammond2021empirical}. Released on June 2021, GitHub Copilot has recently emerged as an ``AI pair programmer'', which is powered by OpenAI Codex and suggests code or entire functions in IDEs as a plug-in \cite{githubcopilot} to help developers achieve code auto-completion in programming activities.

Although the emergence of AI-assisted programming tools has empowered practitioners in their software development efforts, there is little evidence and lack of empirically-rooted studies (e.g, \cite{hammond2021empirical}, \cite{bird2022taking}, \cite{madi2023association}) on the role of AI-assisted programming tools in software development. The existing studies primarily focus on the correctness and understanding of the code suggested by Copilot, and little is known about the practices and challenges of using Copilot with programming activities. \textbf{To close the gap}, we conducted this study that collects data from Stack Overflow (SO) and GitHub Discussions to get practitioners' perspectives on using Copilot during software engineering and development. 

\textbf{The contributions of this work}: (1) we identified the programming languages, IDEs, and technologies used with Copilot; (2) we provided the functions implemented by Copilot, the benefits, limitations, and challenges of using Copilot; and (3) we present the directions to be further explored.



\section{Related Work}
\label{sec:relatedWork}

Several studies focused on the security issues of Copilot. Sandoval \textit{et al.} \cite{gustavo2022security} conducted a user study to investigate the impact of programming with LLMs that support Copilot. Their results show that LLMs have a positive impact on the correctness of functions, and they did not find any decisive impact on the correctness of safety. Several studies focused on the quality of the code generated by Copilot. Imai \cite{imai2022github} compared the effectiveness of programming with Copilot versus human programming, and found that the generated code by Copilot is inferior than human-written code. Yetistiren \textit{et al.} \cite{yetistiren2022assessing} assessed the quality of generated code by Copilot in terms of validity, correctness, and efficiency. Their empirical analysis shows Copilot is a promising tool. Madi \textit{et al.} \cite{madi2023association} focused on readability and visual inspection of Copilot generated code. Through a human experiment, their study highlights that programmers should beware of the code generated by tools. Wang \textit{et al.} \cite{wang2023practitioners} collected practitioners' expectations on code generation tools through a mixed-methods approach. They found that effectiveness and code quality is more important than other expectations. Several studies focused on the limitations and challenges in Copilot assisted programming. Dakhel \textit{et al.} \cite{dakhel2022asset} explored the capabilities of Copilot through empirical evaluations, and their results suggest that Copilot shows limitations as an assistant for developers. Nguyen and Nadi \cite{nguyen2022evalution} conducted an empirical study to evaluate the correctness and comprehensibility of the code suggested by Copilot. Their findings revealed that Copilot’s suggestions for different programming languages do not differ significantly, and they identified potential shortcomings of Copilot, like generating complex code. Bird \textit{et al.} \cite{bird2022taking} conducted three studies to understand how developers use Copilot and their findings indicated that developers spent more time assessing Copilot's suggestions than doing the task by themselves. Sarkar \textit{et al.} \cite{sarkar2022artificial} compared programming with Copilot to previous conceptualizations of programmer assistance to examine their similarities and differences, and discussed the issues that might arise in applying LLMs to programming.

Compared to the existing work (e.g., \cite{yetistiren2022assessing}, \cite{dakhel2022asset}), our work intends to understand the practices and challenges of Copilot by exploring the programming languages, IDEs, technologies used with Copilot, functions implemented by Copilot, and the benefits, limitations, and challenges of using Copilot.

\section{Research Design}
\label{sec:researchDesign}


\subsection{Research Questions}
\label{subsec:researchQuestions}
\begin{figure}[htbp]
	\centering
	\includegraphics[width=1.0\linewidth]{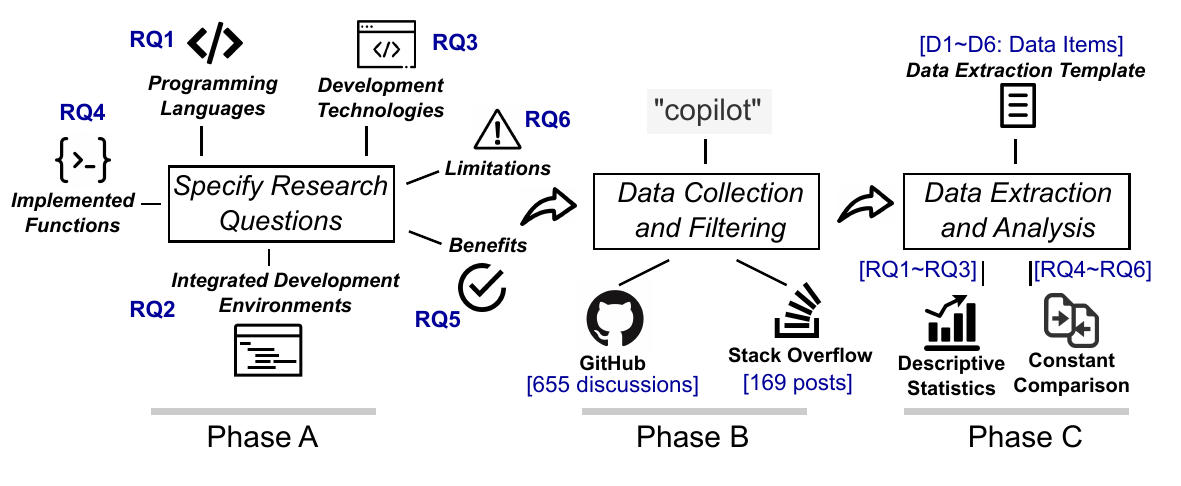}
	\caption{Overview of the research process}
	\label{fig:Overview of research process}
\end{figure}

The Research Questions (RQs) and their rationale are presented in Table \ref{Research Questions and their Rationales}. The overview of the research process is shown in Figure \ref{fig:Overview of research process} and detailed below.

\begin{table}[htbp]
\scriptsize
\caption{Research Questions and their Rationales}
\label{Research Questions and their Rationales}
\begin{tabular}{p{2.5cm}p{5.55cm}}
\hline 
\textbf{Research Question}                                                 & \textbf{Rationale}                                     \\ \hline
\textbf{RQ1}: What programming languages are used with GitHub Copilot?     & Copilot can help practitioners write less code. This RQ aims to collect the programming languages that developers tend to use with Copilot.                  \\ \hline   
\textbf{RQ2}: What IDEs are used with GitHub Copilot?                 & Copilot is a third-party plug-in used in IDEs. This RQ aims to identify the IDEs frequently used with Copilot. The answers of this RQ can help developers choose which IDE to use when they code with Copilot.     \\ \hline
\textbf{RQ3}: What technologies are used with GitHub Copilot?              & When writing code, programmers need to employ certain technologies to complete the development. This RQ aims to investigate the technologies that can be used with Copilot (e.g., frameworks), and the answers of this RQ can help developers to choose the technologies when they use Copilot.                                                                                                                       \\ \hline
\textbf{RQ4}: What functions are implemented by using GitHub Copilot?      & Copilot can complete entire functions according to users' comments. This RQ aims to provide a categorization of the functions that can be implemented by Copilot, and the answers of this RQ can provide developers guidance when implementing functions by using Copilot. \\ \hline
\textbf{RQ5}: What are the benefits of using GitHub Copilot?               & Using Copilot to assist programming can bring many benefits (e.g., reducing the workload of developers). This RQ aims to collect the advantages brought to the development by applying Copilot.                                                      \\ \hline
\textbf{RQ6}: What are the limitations and challenges of using GitHub Copilot?   & Although using Copilot to assist in writing code can help developers with their programming activities, there are still restrictions and problems when using Copilot. This RQ aims to collect and identify the limitations and challenges practitioners may experience when using Copilot. The answers of this RQ can help practitioners make an informed decision when deciding whether to code with the help of Copilot.   \\ \hline
\end{tabular}
\end{table}

\subsection{Data Collection and Filtering}
This study focuses on understanding the practices and challenges of using Copilot collected from SO and GitHub Discussions. SO is a popular software development community and has been widely used by developers to ask and answer questions as a Q\&A platform. GitHub Discussions is a feature of GitHub used to support the communication among the members of a project. Different from SO, GitHub Discussions can provide various communication intentions, not just question-answering (e.g., a discussion can report errors or discuss the potential development of a software project) \cite{hata2022github}, from which the data can be complementary to the data from SO. Besides, GitHub Discussions can provide a center of a community knowledge base connected with other artifacts in a project \cite{hata2022github}. Therefore, we decided to use SO and GitHub Discussions as the data sources to answer our RQs, and we conducted the searches for both SO and GitHub Discussions on November 23rd, 2022. 

\textbf{For SO}, ``\textit{copilot}'' is used as the term to search the posts related to Copilot. After searching, we got a total of 557 posts that include the search term ``\textit{copilot}''. The term ``\textit{copilot}'' may appears more than once in a post, so there may be duplicates in the URL collection of these retrieved posts. After removing the duplicates, we ended up with 521 posts with unique URLs. To manually label posts related to Copilot, we conducted a pilot data labelling by two authors with 10 retrieved SO posts. Specifically, the inclusion criterion is that the post must provide information referring to Copilot. We calculated the Cohen's Kappa coefficient \cite{jacob1960coefficient} to measure the consistency of labelled posts, which is 0.773, thus indicating a decent agreement between the two coders. After excluding the irrelevant posts in the search results, we finally got 169 Copilot related SO posts.

\textbf{For GitHub Discussions}, GitHub discussions are organized according to categories. After exploring the categories on GitHub Discussions, we found the ``Copilot'' category which contains the feedback, questions, and conversations about Copilot \cite{githubdisucssions} under the ``GitHub product categories''. Since all the discussions under the ``Copilot'' category are related to Copilot, we then included all the discussions under the ``Copilot'' category as related discussions to extract data. The number of discussions related to Copilot is 655.

\subsection{Data Extraction and Analysis} 
\subsubsection{Extract Data}
To answer the RQs in Section \ref{subsec:researchQuestions}, we extracted the data items listed in Table \ref{Data items extracted with their corresponding RQs and analysis methods}. The first and third authors conducted a pilot data extraction independently with 10 SO posts and 10 GitHub discussions randomly selected from the 169 SO posts and 655 GitHub discussions. The second author was involved to discuss with the two authors and came to an agreement if any disagreements were found during the pilot. After the pilot, the criteria for data extraction were determined: (1) for all the data items listed in Table \ref{Data items extracted with their corresponding RQs and analysis methods}, they will be extracted and counted only if they are explicitly mentioned by developers that they were used with Copilot; (2) if the same developer repeatedly mentioned the same data item in an SO post or a GitHub discussion, we only counted it once. In a post or discussion, multiple developers may mention Copilot related data items, resulting in situation that the total number of instances of certain data item extracted may be greater than the number of posts and discussions. The first and third authors further extracted the data items from the filtered posts and discussions according to the extraction criteria, marked uncertain parts, and discussed with the second author to reach a consensus. Finally, the first author rechecked all the extraction results by the two authors from the filtered posts and discussions to ensure the correctness of the extracted data.

\begin{table}[htbp]
\scriptsize
\caption{Data items extracted with their corresponding RQs and analysis methods}
\label{Data items extracted with their corresponding RQs and analysis methods}
\begin{tabular}{p{0.15cm}<{\centering}p{1.6cm}p{3.2cm}p{1.4cm}p{0.3cm}}
\hline
\textbf{\#} & \textbf{Data Item}                     & \textbf{Description}                                             & \textbf{Analysis Method}         & \textbf{RQ}   \\ \hline
D1          & Programming language                   & \textit{Programming language used with Copilot}                  & Descriptive statistics \cite{christopher2017interpreting}      & RQ1    \\ \hline
D2          & IDE                                    & \textit{IDEs used with Copilot}                                  & Descriptive statistics           & RQ2    \\ \hline
D3          & Technology                             & \textit{Technologies used with Copilot}                          & Descriptive statistics                & RQ3    \\ \hline
D4          & Function                               & \textit{Functions implemented by Copilot}                        & Constant comparison                   & RQ4    \\ \hline
D5          & Benefit                                & \textit{Benefits brought by using Copilot}                       & Constant comparison                   & RQ5    \\ \hline
D6          & Limitation and Challenge               & \textit{The restrictions and difficulties when using Copilot}    & Constant comparison                   & RQ6    \\ \hline
\end{tabular}
\end{table}

\subsubsection{Analyze Data}
For RQ1, RQ2, and RQ3, we used descriptive statistics \cite{christopher2017interpreting} to analyze and present the results. For RQ4, RQ5, and RQ6, we conducted a qualitative data analysis by applying the Constant Comparison method \cite{glaser1965the}. We constantly compared each part of the data (e.g., emergent codes) to explore differences and similarities in the extracted data to form categories \cite{core2006lillemor}. Note that for answering RQ4, we categorized the functions (D4) based on developers' discussions, i.e., developers' descriptions of the mentioned functions. Firstly, the first and the third authors coded the filtered posts and discussions with the corresponding data items listed (see Table \ref{Data items extracted with their corresponding RQs and analysis methods}). Secondly, the first author reviewed the coded data by the third author to make sure the extracted data were coded correctly. Finally, the first author combined all the codes into higher-level concepts and turned them into categories. After that, the second author examined the coding and categorization results, in which any divergence was discussed till the three authors reached an agreement. The data analysis methods with their corresponding data items and RQs are listed in Table \ref{Data items extracted with their corresponding RQs and analysis methods}. The data analysis results are provided in \cite{replpack}.


\section{Results}
\label{sec:results}
\begin{figure*}[htbp]
	\centering
	\includegraphics[width=0.68\linewidth]{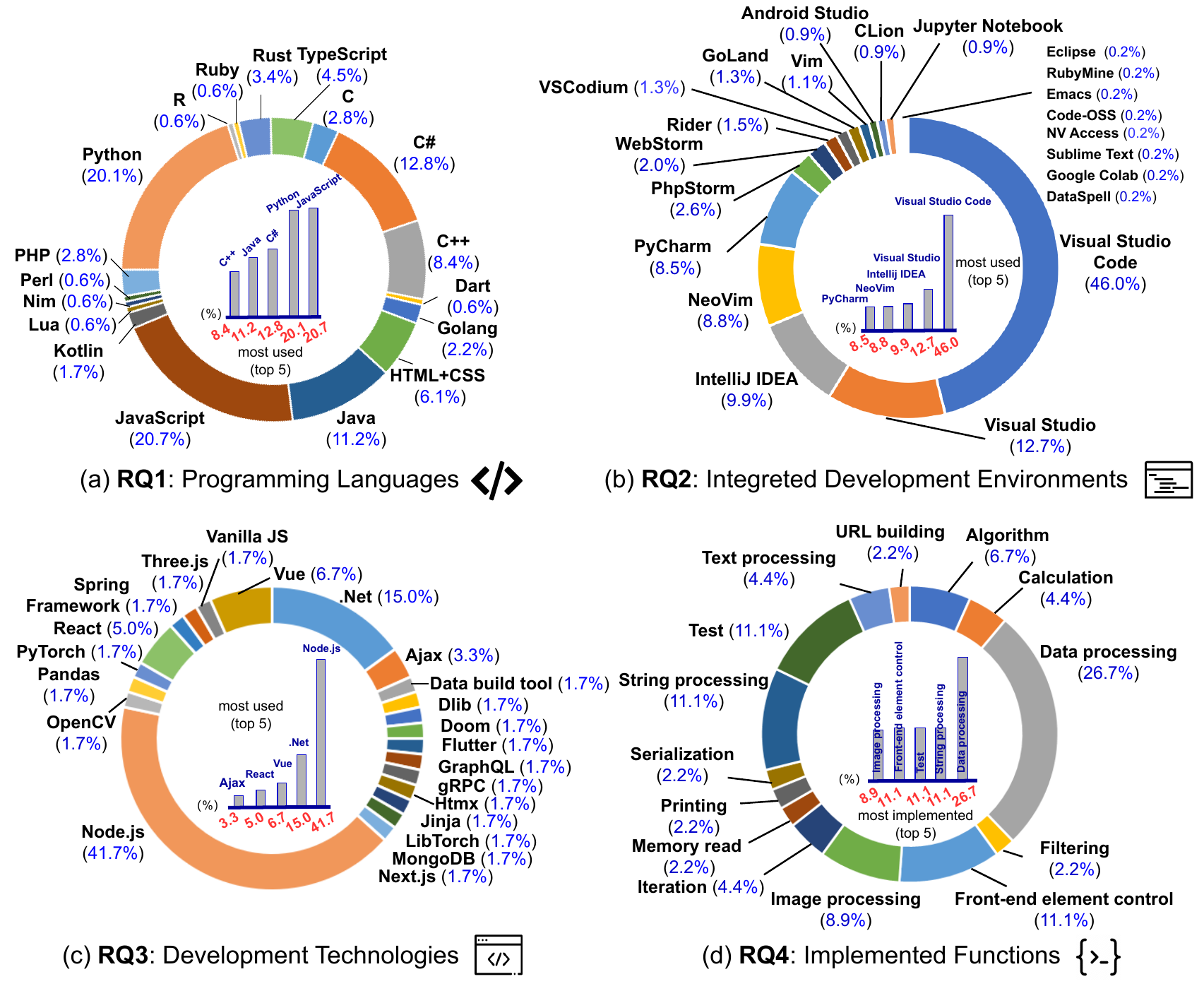}
	\caption{Programming languages, IDEs, technologies, and implemented functions of using Copilot (results of RQ1 to RQ4)}
	\label{fig:Results of RQ1-RQ4}
\end{figure*}

This section presents the results of the study, in which the results of RQ1 to RQ4 are visualized in Fig \ref{fig:Results of RQ1-RQ4}, and the results of RQ5 and RQ6 are provided in Table \ref{Benefits of using Copilot} and \ref{Limitations and challenges of using Copilot}.

\noindent \textbf{RQ1: Programming languages used with GitHub Copilot}\\
Figure \ref{fig:Results of RQ1-RQ4}a lists the 18 programming languages used with Copilot, in which \textit{JavaScript} and \textit{Python} are the most frequently used ones, both accounting for one fifth. Besides, developers often write \textit{C\#} and \textit{Java} code when using Copilot, as one practitioner mentioned ``\textit{the GitHub Copilot extension is enabled in my VS 2022 C\# environment}'' (GitHub \#14115). \textit{TypeScript}, \textit{Rust}, \textit{PHP}, \textit{C}, \textit{Golang}, and \textit{Kotlin} were used with Copilot between 3\textasciitilde8 times each (1.7\% to 8.4\%). The rest programming languages (e.g., \textit{Perl} and \textit{Ruby}), which are not popular, were mentioned only once with Copilot.

\noindent \textbf{RQ2: IDEs used with GitHub Copilot}\\
Figure \ref{fig:Results of RQ1-RQ4}b shows 22 types of IDEs that are used with Copilot. \textit{Visual Studio Code} is the dominant IDE, accounting for 46.0\%. When first released, Copilot only worked with \textit{Visual Studio Code} editor, and it is expected that \textit{Visual Studio Code} is the IDE most often used with Copilot. Mainstream code editors, including \textit{Visual Studio}, \textit{IntelliJ IDEA}, \textit{NeoVim}, and \textit{PyCharm} are also occasionally used, account for 39.9\% in total. The remaining IDEs were rarely mentioned by developers, and one possible reason is that there are often \textit{integration issues} when using Copilot within them according to the results of RQ6.

\noindent \textbf{RQ3: Technologies used with GitHub Copilot}\\
Figure \ref{fig:Results of RQ1-RQ4}c presents 22 technologies used with Copilot. We find that these identified technologies include frameworks, APIs, and libraries. \textit{Node.js}, whose proportion is more than 40\%, is one of the most popular back-end runtime environments for \textit{JavaScript}, which is also the most frequently used language with Copilot (see the results of RQ1), thus it is reasonable that \textit{Node.js} is the major technology used with Copilot. In addition, \textit{.NET} which works for Web development, and \textit{Vue}, \textit{React}, and \textit{Ajax} which are frameworks for front-end development, were mentioned less often compared to \textit{Node.js}. The rest of the identified technologies, many of which relate to machine learning (e.g., \textit{Pandas}, \textit{Dlib}, and \textit{OpenCV}) or front-end development (e.g., \textit{Htmx}, \textit{Vanilla JS}, and \textit{Next.js)}, are rarely used with Copilot, and each of them appears only once.

\begin{table*}[htbp]
\caption{Benefits of using Copilot (results of RQ5)}
\scriptsize
\label{Benefits of using Copilot}
\begin{tabular}{p{3.25cm}p{12cm}p{0.6cm}<{\centering}p{0.6cm}<{\centering}}
\hline
\textbf{Benefit}                                         & \textbf{Example}        & \textbf{Count}   & \textbf{\%} \\ \hline
Useful code generation                                   & \textit{I find myself writing a lot of tests, and Copilot is excellent at helping with writing repetitive tests} (GitHub \#9282)                                                                            & 24               & 49.0\%      \\ \hline
Faster development                                       & \textit{I really enjoy using it , it reduce programming time} (GitHub \#17382)                                            
                                                                                   & 8                & 16.3\%       \\ \hline
Better code quality                                      & \textit{it's faster and simpler to your solution} (SO \#68418725)                                                                
                                                                                   & 5                & 10.2\%       \\ \hline
Good adaptation to users' code patterns                  & \textit{GitHub copilot adapt to your coding practices} (SO \#69740880)                                                          
                                                                                   & 3                & 6.1\%        \\ \hline
Better user experience                                   & \textit{Since copilot works totally different compared to all the other products out there, it is a lot more fun to use and does not annoy me like some other AI systems} (GitHub \#7254)                    & 3                & 6.1\%        \\ \hline
Powerful code interpretation and conversion functions    & \textit{Does Copilot have the code explanation feature or something similar? It does! some active members were given beta access.} (GitHub \#38089)                                                                                                                      & 2                & 4.1\%        \\ \hline
Frequent updates to provide more features                & \textit{Keep in mind that there are updates to the plugin very frequently, so there's still hope} (SO \#70428218)                                                                                                   & 1                & 2.0\%        \\ \hline
Free for students                                        & \textit{If you are a student you can sign up for the GitHub Student Pack, which gives a lot of benefits, one being copilot for free} (GitHub \#31494)                                                        & 1                & 2.0\%        \\ \hline
Strong integration capability                            & \textit{GitHub is supporting more editors} (GitHub \#6858)                                                                                                                                                    & 1                & 2.0\%        \\ \hline
Ease of study and use                                    & \textit{when using this plugin, can study at a relatively low cost} (GitHub \#8028)                                                                                                                           & 1                & 2.0\%        \\ \hline
\end{tabular}
\end{table*}

\begin{table*}[htbp]
\scriptsize
\caption{Limitations and challenges of using Copilot (results of RQ6)}
\label{Limitations and challenges of using Copilot}
\begin{tabular}{p{3cm}p{12.25cm}p{0.6cm}<{\centering}p{0.6cm}<{\centering}}
\hline
\textbf{Limitation \& Challenge}                   & \textbf{Example}        & \textbf{Count}   & \textbf{\%} \\ \hline
Difficulty of integration                          & \textit{Copilot only works with VSCode, VSCodium is not supported at the moment} (GitHub \#14837)                                                                                                             & 75               & 28.0\%      \\ \hline
Difficulty of accessing Copilot                    & \textit{I cannot connect to the GitHub account and the Copilot server in VSCode, also cannot use the Copilot plugin} (SO \#74398521)                                                                                & 47               & 17.5\%      \\ \hline
Limitation to code generation                      & \textit{Copilot is limited to around 1000 characters in the response} (GitHub \#15122)                                           
                                                                             & 39               & 14.6\%       \\ \hline
Poor quality of generated code                     & \textit{Github Copilot suggest solutions that don't work} (SO \#73701039)                                                             
                                                                             & 31               & 11.6\%       \\ \hline
Code privacy threat                                & \textit{Copilot does collect personal data so just take precaution when working in private repos} (GitHub \#7163)                                                                                             & 20               & 7.5\%       \\ \hline
Unfriendly user experience                         & \textit{I had the same problem today, an amazing tool with poor user experience} (GitHub \#8468)                                                                                                              & 14               & 5.2\%       \\ \hline
High pricing                                       & \textit{it is obvious that no one in South America will pay that price, it is too expensive} (GitHub \#24594)                                                                                                 & 11               & 4.1\%       \\ \hline
Difficulty of understanding the generated code     & \textit{I really do not understand this enough, and have no idea half of what this code does honestly. It was written by Copilot.} (SO \#72282605)                                                                  & 9                & 3.4\%       \\ \hline
No edition for organizations                       & \textit{Currently, Copilot is only available for individual user accounts and organizations aren't able to purchase/manage Copilot for their members just yet} (GitHub \#32775)                          & 7                & 2.7\%       \\ \hline
Lack of customization                              & \textit{My question is about setting up shortcuts in Visual Studio Code VSCode for GitHub Copilot Labs.} (SO \#73564811)                                                                                            & 5                & 1.9\%       \\ \hline
Difficulty of subscription                         & \textit{My copilot subscription suddenly stopped. Tried log out and in. Never have reply on support ticket over 10 days} (GitHub \#36190)                                                                     & 3                & 1.1\%       \\ \hline
Challenge of not providing outdated suggestions    & \textit{making sure that the tool does not provide outdated suggestions would still be a challenge} (SO \#72554382)                                                                                                 & 2                & 0.7\%       \\ \hline
Show loading                                       & \textit{I am not sure what is causing this but while editing files within Visual Studio, I am periodically locking up with the following dialog showing} (SO \#73682137)                                         & 2                & 0.7\%       \\ \hline
Hard to configure                                  & \textit{Keep getting "Your Copilot experience is not fully configured, complete your setup" in Visual Studio 2022} (GitHub \#19556)                                                                     & 2                & 0.7\%       \\ \hline
Need of basic programming knowledge                & \textit{It is useless if you do not understand the programming language or the task you want to do} (GitHub \#35850)                                                                                          & 1                & 0.4\%       \\ \hline
\end{tabular}
\end{table*}

\noindent \textbf{RQ4: Functions implemented by using GitHub Copilot}\\
Figure \ref{fig:Results of RQ1-RQ4}d shows 14 functions implemented by using Copilot. The main function implemented by Copilot is \textit{data processing}, indicating that developers tend to make use of Copilot to write functions working with data. Besides, \textit{Front-end element control}, \textit{string processing}, and \textit{Test} account for the same, i.e., 11.1\%. When implementing functions, developers also use Copilot to code \textit{image processing}, \textit{algorithm}, \textit{iteration}, \textit{calculation}, \textit{filtering}, \textit{printing}, \textit{memory read}, \textit{serialization}, and \textit{URL building}, which range from 2.2\% to 8.9\%.


\noindent \textbf{RQ5: Benefits of using GitHub Copilot}\\
Table \ref{Benefits of using Copilot} highlights 10 benefits of using Copilot. Most developers mentioned that they used Copilot for \textit{useful code generation}, which reduced their workload and gave them help when they have no idea about how to write code. Programming with Copilot also brings \textit{faster development}, as one discussion remarked, Copilot ``\textit{saves developers a lot of time}'' (GitHub \#35850). Meanwhile, \textit{better code quality} can be obtained by using Copilot. Compared to the code written by developers themselves, the code suggested by Copilot is usually shorter and more correct, as one developer said, ``\textit{often Copilot is smarter than me}'' (SO \#74512186). Copilot can use machine learning models to learn code style of developers, so as to offer \textit{good adaptation to users' code patterns}. Three developers mentioned that Copilot can give them \textit{better user experience} than other AI-assisted programming tools, for example, one developer stated that ``\textit{Copilot works totally different compared to all the other products out there, it is a lot more fun to use and does not annoy me like some other AI systems}'' (GitHub \#7254), without providing the names of the other products.

\noindent \textbf{RQ6: Limitations and challenges of using GitHub Copilot}\\
Table \ref{Limitations and challenges of using Copilot} lists 15 limitations and challenges of using Copilot. Most developers pointed out the \textit{difficulty of integration} between Copilot and IDEs or other plug-ins. After Copilot was installed in developers' IDEs, certain plug-ins did not work and Copilot may conflict with some shortcut settings of the editors. Moreover, Copilot cannot be successfully integrated with some IDEs as Copilot does not support these editors yet. Due to the instability of Copilot servers, developers may have \textit{difficulties of accessing Copilot}. The code suggested by Copilot has restrictions as well, and sometimes it just offers few solutions, which are not enough for users, which brings \textit{limitation to code generation}, as one developer said ``\textit{multiple solution is too little}'' (GitHub \#37304). Practitioners also complained about the \textit{poor quality of generated code} by Copilot. Some practitioners said that ``\textit{GitHub Copilot suggest solutions that don't work}'' (SO \#73701039), and some practitioners found that when the code files became larger, the quality of the code suggested by Copilot ``\textit{becomes unacceptable}'' (GitHub \#9282). When using Copilot, developers pay much attention to \textit{code privacy threat} as well. They were worried that Copilot may use their code information without permission. Contrary to the developers who mentioned that Copilot gave them a \textit{better user experience} than other AI-assisted programming tools, some practitioners said they had an \textit{unfriendly user experience} when coding with Copilot.

\section{Implications}
\label{sec:implications}

\textbf{Integration of Copilot with IDEs}:
According to the results of RQ2 and RQ6, we found that most developers choose to integrate the Copilot plug-in in mainstream IDEs (including \textit{Visual Studio Code}, \textit{Visual Studio}, \textit{IntelliJ IDEA}, \textit{NeoVim}, and \textit{PyCharm}), and the percentage of mainstream IDEs used with Copilot by practitioners reaches 85.9\%. When developers choose the lesser known IDEs (e.g., \textit{Sublime Text}), they often find it hard to integrate the Copilot plug-in and thus have \textit{difficulty of integration}. In addition to the reason that developers may install Copilot in their chosen IDEs incorrectly, another reason for the \textit{difficulty of integration} is that Copilot does not support certain IDEs at the moment. When developers choose to use Copilot in mainstream IDEs, they can install it smoothly, and even if problems arise during the installation or use, they can easily find a solution on SO or GitHub Discussions as many other developers may have encountered similar issues. To reduce the \textit{difficulty of integration}, we recommend practitioners to use mainstream IDEs with Copilot. Besides, Copilot may consider improving the integration of Copilot by supporting more IDEs in the future.

\textbf{Support for Front-end and Machine Learning Development}:
As we can see from the results of RQ1, RQ3, and RQ4, practitioners often write \textit{JavaScript} and \textit{Python} code when using Copilot, and they tend to use Copilot with front-end and machine learning related technologies (including frameworks, APIs, and libraries) to implement front-end (e.g., \textit{front-end element control}) and machine learning functions (e.g., \textit{data processing} and \textit{image processing}). \textit{JavaScript} is the foundation language of many popular front-end frameworks and most of Websites use \textit{JavaScript} on the client side. \textit{Python} is the first choice when it comes to the development of machine learning solutions with the help of rich libraries, e.g., \textit{OpenCV}. It is consequently reasonable that developers tend to use Copilot with \textit{JavaScript} to facilitate and generate code for front-end and \textit{Python} for machine learning development.


\textbf{Potentials and Perils of Using Copilot in Software Development}: Trained on billions of lines of code, Copilot can turn natural language prompts into coding suggestions across dozens of programming languages and make developers code faster and easier \cite{githubcopilot}. The results of RQ5 and RQ6 show that many benefits of using Copilot contradict its limitations and challenges, e.g., \textit{useful code generation} vs. \textit{limitation to code generation}. When deciding to use Copilot, developers should consider tool integration, user experience, budget, code privacy, and some other aspects, and make trade-offs between these factors. In short, using Copilot is like a double-edged sword, and practitioners need to consider various aspects carefully when deciding whether or not to use it. If Copilot can be used with appropriate programming languages and technologies to implement functions required by users correctly in developers' IDEs, it will certainly optimize developers' coding workflow and do what matters most - building software by letting AI do the redundant work. Otherwise, it will bring difficulties and restrictions to development, making developers feel frustrated and constrained. The study results can help practitioners being aware of the potential advantages and disadvantages of using Copilot and thus make an informed decision whether to use it for programming activities.

\textbf{Towards an Effective Use of Copilot}: Further investigation about the practices of Copilot can be conducted by questionnaire and interview. Under what conditions the challenges of using Copilot will show up as advantages or disadvantages, and how to use Copilot to convert its disadvantages into advantages are also worth further exploration. Besides, although we have investigated various aspects of using Copilot (e.g., limitations and challenges), we have not looked in depth at what types of users (e.g., developers, educators, and students) who use Copilot, when and how they use Copilot, and for what specific purposes. By exploring these aspects, researchers can get meaningful information which would help guide towards an effective use of Copilot.

\section{Threats to Validity}
\label{sec:threats}

\textbf{Construct validity}: We conducted data labelling, extraction, and analysis manually, which may lead to personal bias. To reduce this threat, the data labelling of SO posts was performed after the pilot labelling to reach an agreement between the authors. The data extraction and analysis was also conducted by two authors, and the first author rechecked all the results produced by the third author. During the whole process, the first author continuously consulted with the second author to ensure there are no divergences.

\textbf{External validity}: We chose two popular developer communities (SO and GitHub Discussions) because SO has been widely used in software engineering studies and GitHub Discussions is a new feature of GitHub for discussing specific topics \cite{hata2022github}. These two data sources can partially alleviate the threat to external validity. However, we admit that our selected data sources may not be representative enough to understand all the practices and challenges of using Copilot.

\textbf{Reliability}: We conducted a pilot labelling before the formal labelling of SO posts with two authors, and the Cohen's Kappa coefficient is 0.773, indicating a decent consistency. We acknowledge that this threat might still exist due to the small number of posts used in the pilot. All the steps in our study, including manual labelling, extraction, and analysis of data were conducted by three authors. During the process, the three authors discussed the results until there was no any disagreements in order to produce consistent results. In addition, the dataset of this study that contains all the extracted data and labelling results from the SO posts and GitHub discussions has been provided online for validation and replication purposes \cite{replpack}.

\section{Conclusions}
\label{sec:conclusions}
We conducted an empirical study on SO and GitHub Discussions to understand the practices and challenges of using GitHub Copilot from the practitioners' perspective. We used ``\textit{copilot}'' as the search term to collect data from SO and collected all the discussions under the ``Copilot'' category in GitHub Discussions. Finally, we got 169 SO posts and 655 GitHub discussions related to Copilot. Our results identified the programming languages, IDEs, technologies used with Copilot, functions implemented by Copilot, and the benefits, limitations, and challenges of using Copilot, which are first-hand information for developers. 

In the next step, we plan to further explore when to use Copilot, for what specific purposes, and by whom, which helps to guide towards an effective use of Copilot. 


\balance
\bibliographystyle{IEEEtran}
\bibliography{References}

\end{document}